\pdfoutput=1																				
\documentclass[
	12pt,																						
	twoside,																					
	headsepline=0.4pt																
	]{scrartcl}																			
\usepackage[utf8]{inputenc}													
\usepackage[T1]{fontenc}														
\usepackage{lmodern}																
\usepackage[british]{babel}													
\usepackage{ragged2e}																
\usepackage{microtype}															
\usepackage{mparhack}																
\usepackage[dvipsnames,hyperref]{xcolor}							
\usepackage{graphicx}																
\usepackage{pdfpages}																
\usepackage[
	justification=justified,													
	singlelinecheck=false,														
	font=small,																			
	labelfont=bf,																		
	format=plain,																		
	indention=3mm																		
	]{caption}																				
\usepackage{afterpage}															
\usepackage{dpfloat}																
\usepackage[centertags]{amsmath}										
\usepackage{amssymb}																
\usepackage{isotope}																
\usepackage{upgreek}																
\usepackage{textcomp}																
\usepackage{url}																		
\usepackage[square,comma,numbers]{natbib}						
\usepackage[
	colorlinks=true,																	
	urlcolor= MidnightBlue,														
	urlbordercolor=red,																
	linkcolor=black,																	
	citecolor=MidnightBlue,														
	linkbordercolor=lightgray,												
	breaklinks=true,																	
	hyperfootnotes=true,															
	pdfpagelabels,																		
	pdfstartview=Fit,																
	pdfview=Fit,																			
	bookmarks=true,																	
	bookmarksnumbered,																
	pdfpagemode=UseOutlines,													
	bookmarksopen=false,															
	bookmarksopenlevel=1,															
	pdfdisplaydoctitle=true,													
	pdftitle={Neutrino Telescopes},										
	pdfauthor={Gisela Anton},													
	pdfcreator={LaTeX, hyperref, KOMA-Script},					
	hypertexnames=false																
	]{hyperref}																			
\usepackage{doi}																		
\usepackage[headsepline]{scrlayer-scrpage}						
\usepackage{ftnxtra}																
\usepackage{tablefootnote,footnotehyper}							
\usepackage{etoolbox}																
	\gappto{\UrlBreaks}{\UrlOrds}											
\usepackage{lineno}																	
\usepackage[a4paper]{geometry}											
\newcommand{\unitkm}{\,\text{km}}										
\newcommand{\unitm}{\,\text{m}}											
\newcommand{\unitcm}{\,\text{cm}}										
\newcommand{\unitmm}{\,\text{mm}}										
\newcommand{\unitnm}{\,\text{nm}}										
\newcommand{\unitps}{\,\text{ps}}										
\newcommand{\unitkHz}{\,\text{kHz}}									
\newcommand{\unitmcb}{\,\text{m}^3}									
\newcommand{\unitkmcb}{\,\text{km}^3}								
\newcommand{\unitGeV}{\,\text{GeV}}									
\newcommand{\unitTeV}{\,\text{TeV}}									
\newcommand{\unitPeV}{\,\text{PeV}}									
\newcommand{\unitdeg}{\,^{\circ}}										
\begin{document}

\setcounter{page}{1}										
\pagenumbering{arabic}									
\pagestyle{scrheadings}									
\lehead{\thepage}											
\rohead{\thepage}											
\lefoot{}															
\rofoot{}															
\markleft{G.~Anton}											
\markright{Neutrino Telescopes}					
	\rehead{\leftmark}										
	\lohead{\rightmark}										
\renewcommand*{\headfont}{\normalfont}		

\begin{center}
	\textbf{ \Large{Neutrino Telescopes} }\\
	\vspace{10mm}
	Gisela Anton\\[3mm]
	\textit{Erlangen Centre for Astroparticle Physics,\\ University Erlangen-Nürnberg, \\ Erwin--Rommel-Str.~1, 91058 Erlangen, Germany\\[3mm]
	gisela.anton@fau.de}
\end{center}

\begin{quote}
	This paper addresses the working principle of neutrino telescopes, important detector parameters as well as the layout and performance of current and future neutrino telescopes. It was prepared for the book "Probing Particle Physics with Neutrino Telescopes", C.~P{\'e}rez de los Heros, editor, 2020 (World Scientific) in 2018.
\end{quote}

\thispagestyle{plain}

\section{Introductory considerations}\label{ra_sec1}

Usual astronomical telescopes determine the intensity respectively the spectral density of light arriving from a certain direction at Earth. The analogue task of a neutrino telescope is the detection of neutrinos determining their direction and energy. While visible photons can be collected and directed with mirrors no analogue mechanism exists for neutrinos. This implies that the phase space (area and direction) of neutrinos cannot be changed. So, the strategy is to detect single neutrinos and for each single neutrino to determine its direction and energy from the interaction characteristics and from energy and momenta of secondary particles produced in the neutrino interaction.

Neutrinos only feel the weak force resulting in a rather low probability of interaction with matter. For a large energy regime the neutrino-nucleon cross section increases linearly with energy $E$ while for most of the expected neutrino sources the emitted flux decreases with energy as $E^{-2}$ or even steeper. This implies that a neutrino telescope needs the more target mass the higher the envisaged neutrino energy regime is. To give a rough orientation, usual neutrino detectors employ a Megaton target mass for the GeV energy regime and a Gigaton mass for the TeV energy regime. For more details on neutrino cross sections  and for more details on different neutrino sources and the related fluxes see for instance~\cite{perez20}.

\begin{figure}[t]
	\centering
	\includegraphics[width=0.88\textwidth]{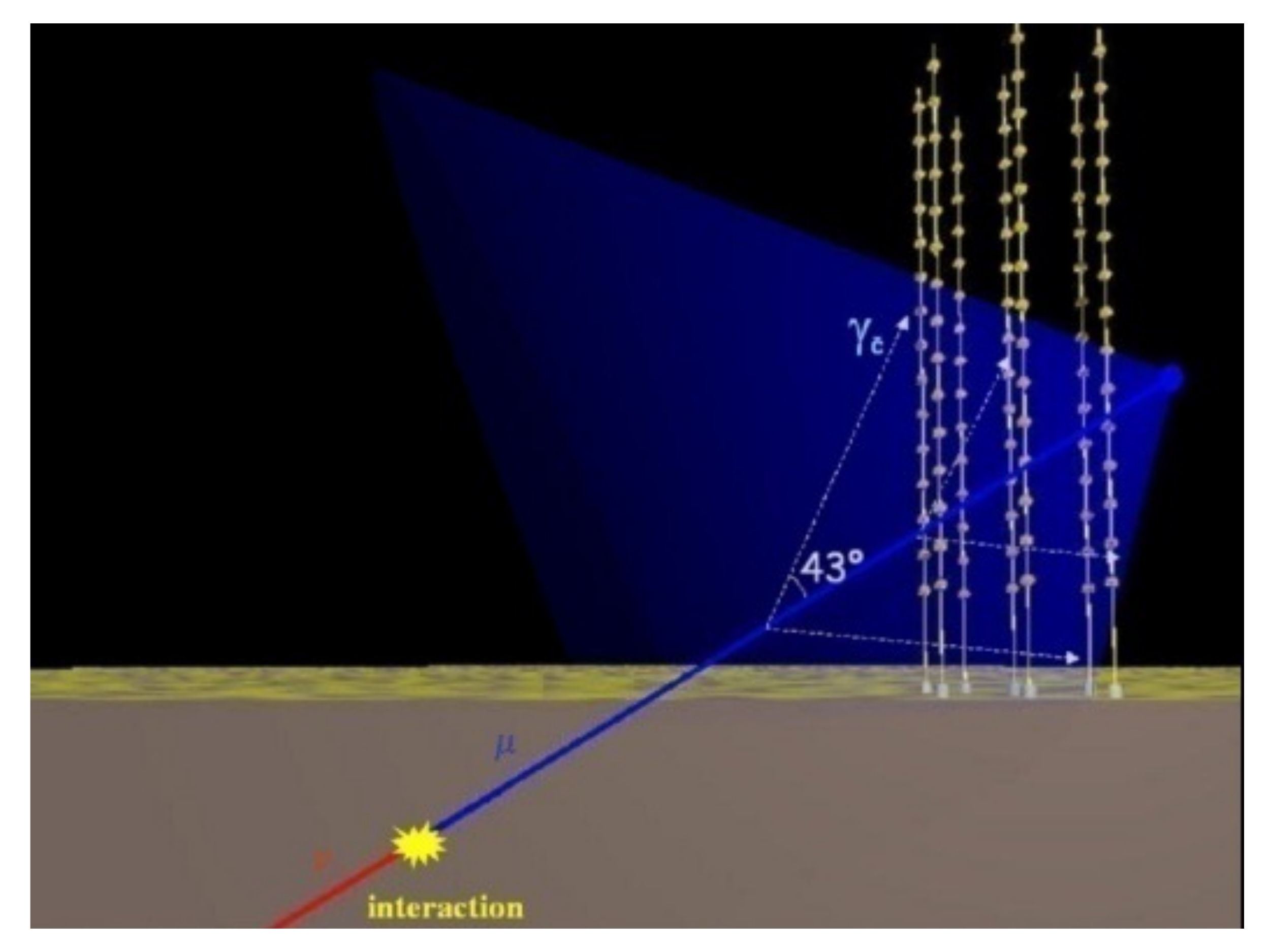}
	\caption{Schematics of neutrino detection in a water Cherenkov telescope: A neutrino (red line) interacts with a nucleus producing a muon (blue line). The muon induces the radiation of Cherenkov photons (blue cone) that can be detected by photo sensors (yellow dots). Credit: ANTARES Collaboration.}
	\label{fig-telescope}
\end{figure}

Due to the large target mass, the instrumentation that is needed to read out the signatures which are produced by a neutrino interacting in the target medium has to be kept simple and cost effective. The established method employs the detection of Cherenkov light that is induced by the charged secondary particles resulting from a neutrino interaction. Fig.~\ref{fig-telescope} shows the example of a muon neutrino $\upnu_{\upmu}$ that interacts with a nucleon producing a muon. The muon travels straight through the target material (for example water) emitting Cherenkov photons. Some of these photons are detected by photo sensors deployed in the target. The trajectory of the muon can be determined from the measured position and time of detected photons. Knowing the probability distribution of longitudinal and transverse momentum transfer from the parent neutrino to the muon the energy and direction of the neutrino can be estimated. According to this detection principle, neutrino telescopes are superior to optical telescopes with respect to the field of view because a neutrino telescope is sensitive to all directions at any time. This property may be advantageous especially for the observation of flaring sources. 

On the other hand, given the low neutrino detection probability it is a draw back that no mechanism exists which can enhance the neutrino sensitivity at the expense of angular field of view. 

In the following two sections we will look a bit deeper into the signatures of neutrinos in a Cherenkov detector.

\section{Interaction signatures of neutrinos}\label{ra_sec2}

\begin{figure}[t]
	\centering
	\includegraphics[width=0.90\textwidth]{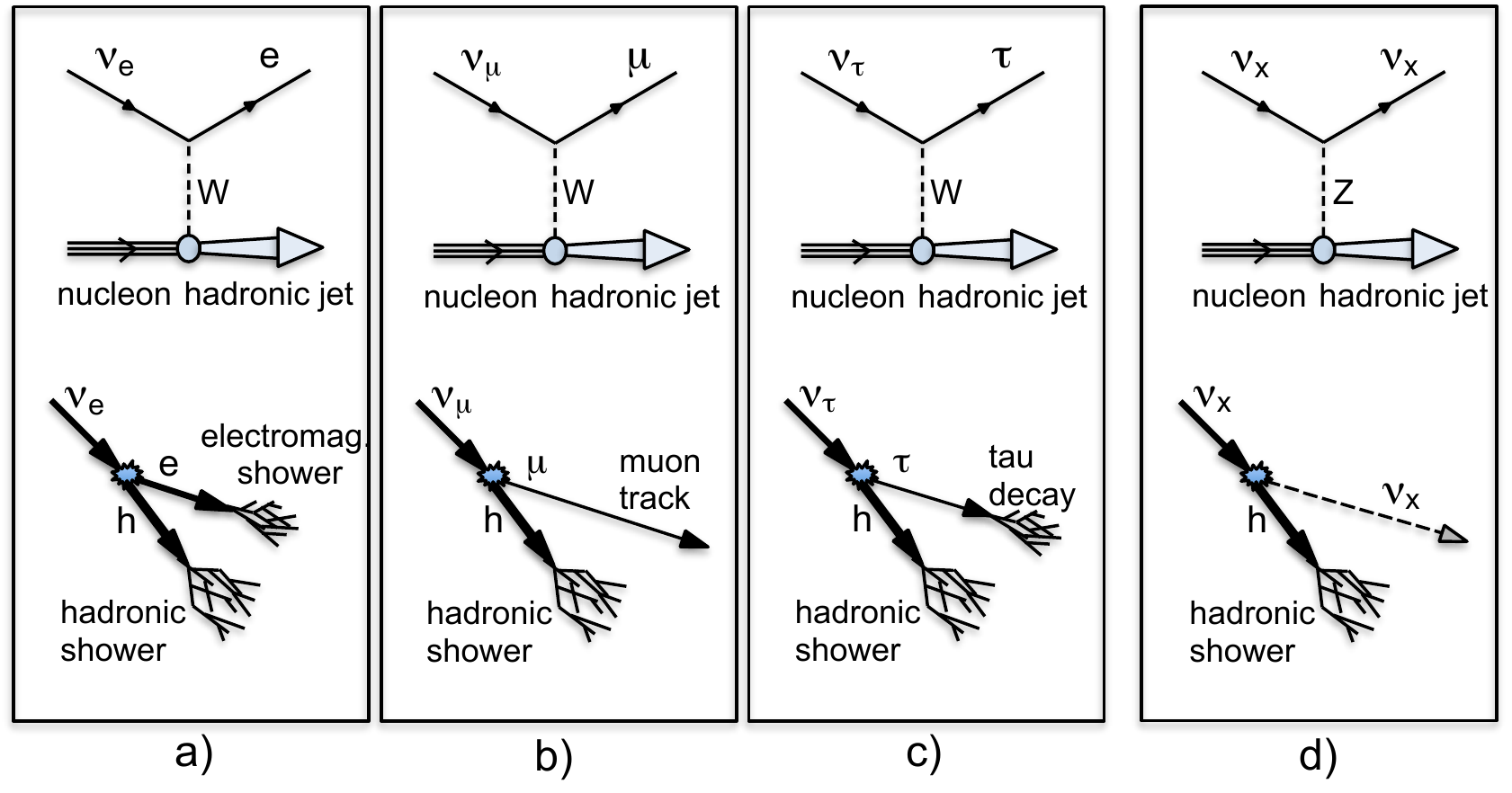}
	\caption{Different types of neutrino interactions with matter. Top: interaction diagrams, bottom: event signatures. Further explanations are given in the text.}
	\label{fig-interaction}
\end{figure}

We consider the neutrino energy regime extending the six orders of magnitude from a few GeV up to a few PeV. Here, the following interaction processes have to be taken into account:

\begin{itemize}
 \item $ \upnu_{\text{e}} \ \text{N} \rightarrow\ \text{e} \ \text{H} $ \\ 
 charged current interaction of an electron neutrino $\upnu_{\text{e}}$ with a nucleon $\text{N}$ producing an electron $\text{e}$ and a hadronic shower $\text{H}$ (see Fig.~\ref{fig-interaction}a)
 \item $ \upnu_{\upmu} \ \text{N} \rightarrow\ \upmu \ \text{H} $ \\ 
 charged current interaction of a muon neutrino $\nu_{\mu}$ with a nucleon $\text{N}$ producing a muon $\upmu$ and a hadronic shower $\text{H}$ (see Fig.~\ref{fig-interaction}b)
 \item $ \upnu_{\uptau} \ \text{N} \rightarrow\ \uptau \ \text{H} $ \\ 
 charged current interaction of a tau neutrino $\nu_{\tau}$ with a nucleon $\text{N}$ producing a tau $\tau$ and a hadronic shower $\text{H}$ (see Fig.~\ref{fig-interaction}c)
 \item $ \upnu_{x} \ \text{N} \rightarrow\ \upnu_{x} \ \text{H} $ \\ 
 neutral current interaction of a neutrino $ \nu_{x} $ of flavour x with a nucleon $\text{N}$ producing a scattered neutrino $\nu_{x}$ and a hadronic shower $\text{H}$ (see Fig.~\ref{fig-interaction}d)
\end{itemize}

The first three processes are so-called charged current interaction processes where a neutrino produces its related charged lepton through W-boson exchange with the nucleon. The kinematics for the three reactions are the same up to the fact that the masses of the electron, muon and tau are different. A considerable part of the neutrino energy is transferred to the charged lepton. The energy that is transferred to the nucleon leads to the production of secondary hadrons which build up the hadronic shower. 

The fourth process is the scattering of a neutrino via Z-boson exchange with the nucleon. Again, the energy that is transferred to the nucleon leads to the production of secondary hadrons, which build up the hadronic shower. 

As a result of the neutrino interaction with the target material we thus can have high energetic electrons, muons, taus and hadronic showers in the final state. In case of anti-neutrinos an anti-lepton is produced in the final state. This does not change the signature of the events very much in the considered energy regime. Accordingly, we further do not distinguish between neutrinos and anti-neutrinos. The neutrino telescope can employ the signature of the charged leptons and the hadronic shower to reconstruct the neutrino parameters, i.e.~neutrino flavour, energy and direction.

The neutrino not only can interact with nucleons but also with the electrons in the target material. The total cross section is proportional to the mass of the target particle, which leads to a much larger cross section for the neutrino scattering from nucleons than scattering from electrons. Accordingly, the interaction with electrons can be neglected in many cases. An important exception occurs at the energy of $6.3\unitPeV$, where the interaction of $\bar{\upnu}_{\text{e}}$ with electrons can proceed via an on-mass-shell W-boson leading to a pronounced resonance peak in the cross section.

The interaction of neutrinos with electrons is of special interest when considering oscillation effects of neutrinos travelling through matter, in our case atmospheric neutrinos travelling through the Earth. While electron neutrinos $\upnu_{\text{e}}$ can be scattered elastically from the electrons via W-boson and Z-boson exchange muon neutrinos $\upnu_{\upmu}$ and tau neutrinos $ \upnu_{\uptau} $ can be scattered elastically via Z--boson exchange only. This mechanism produces differences for the propagation of $\upnu_{\text{e}}$ compared to $\upnu_{\upmu}$ and $\upnu_{\uptau}$ leading to different oscillation patterns depending on normal resp.~inverted mass ordering of the neutrinos~\cite{perez20}.

\section{Signatures of secondary particles in the telescope}\label{ra_sec3}

We now further restrict our assumptions for the neutrino telescope and consider water or ice as target medium. Due to the Cherenkov effect a charged particle travelling with velocity $v = \beta c$ through the medium emits light under an angle $\theta$ relative to the particle momentum following the relation $\cos\theta = 1/\beta n$ with $n$ being the refractive index of the medium \cite{cher37}. For particles travelling with the speed of light $v = c$, the Cherenkov emission angle is fixed to $\theta = 42\unitdeg$ for water and $\theta = 41\unitdeg$ for ice for photons of $400\unitnm$ wavelength. The spectral density of emitted light from a charged particle is given by \cite{patr16}:

\begin{equation}
	\frac{\text{d}^2 N}{\text{d}x\,\text{d}\lambda } = \frac{2\pi\alpha }{\lambda^2}\ \left(1-\frac{1}{\beta^2 n^2 (\lambda)}\right)
\end{equation}

where $\alpha$ is the electromagnetic fine structure constant and $\text{d}N$ is the number of emitted photons per unit of length $\text{d}x$ and unit of wavelength $\text{d}\lambda$ for a particle with unit charge and velocity $\beta$ travelling through a medium with refractive index $n(\lambda)$ provided that $\beta = v/c > 1/n(\lambda)$. Obviously, the spectrum decreases with increasing wavelength. Due to absorption of UV-light in water or ice the relevant wavelength regime for a neutrino telescope extends from $300\unitnm$ to $600\unitnm$. For a minimally ionising particle\footnote{I.e.~for a particle with no further radiative energy loss like Bremsstrahlung or emission of delta electrons, which would lead to additional emission of Cherenkov radiation} the Cherenkov effect leads to about 300 emitted photons per cm track length in this wavelength interval.

\subsection{Signature of electrons}

Electrons induce an electromagnetic shower---often called cascade---comprising bremsstrahlung photons and electron-positron pairs. During the evolution of the shower the energy of secondary electrons, positrons and photons decreases exponentially leading to a relatively short shower length with high energy deposition per unit length and related high density of emitted Cherenkov photons per unit length. Accordingly, the shower length in which $90\,\%$ of the total Cherenkov light is emitted is about $4\unitm$ for a $10\unitGeV$ shower and $7\unitm$ for a $1\unitTeV$ shower. In a huge neutrino telescope this length is almost negligible so that the Cherenkov photons seem to be emitted in a point like process.

\begin{figure}[t]
	\centering
	\includegraphics[width=0.8\textwidth]{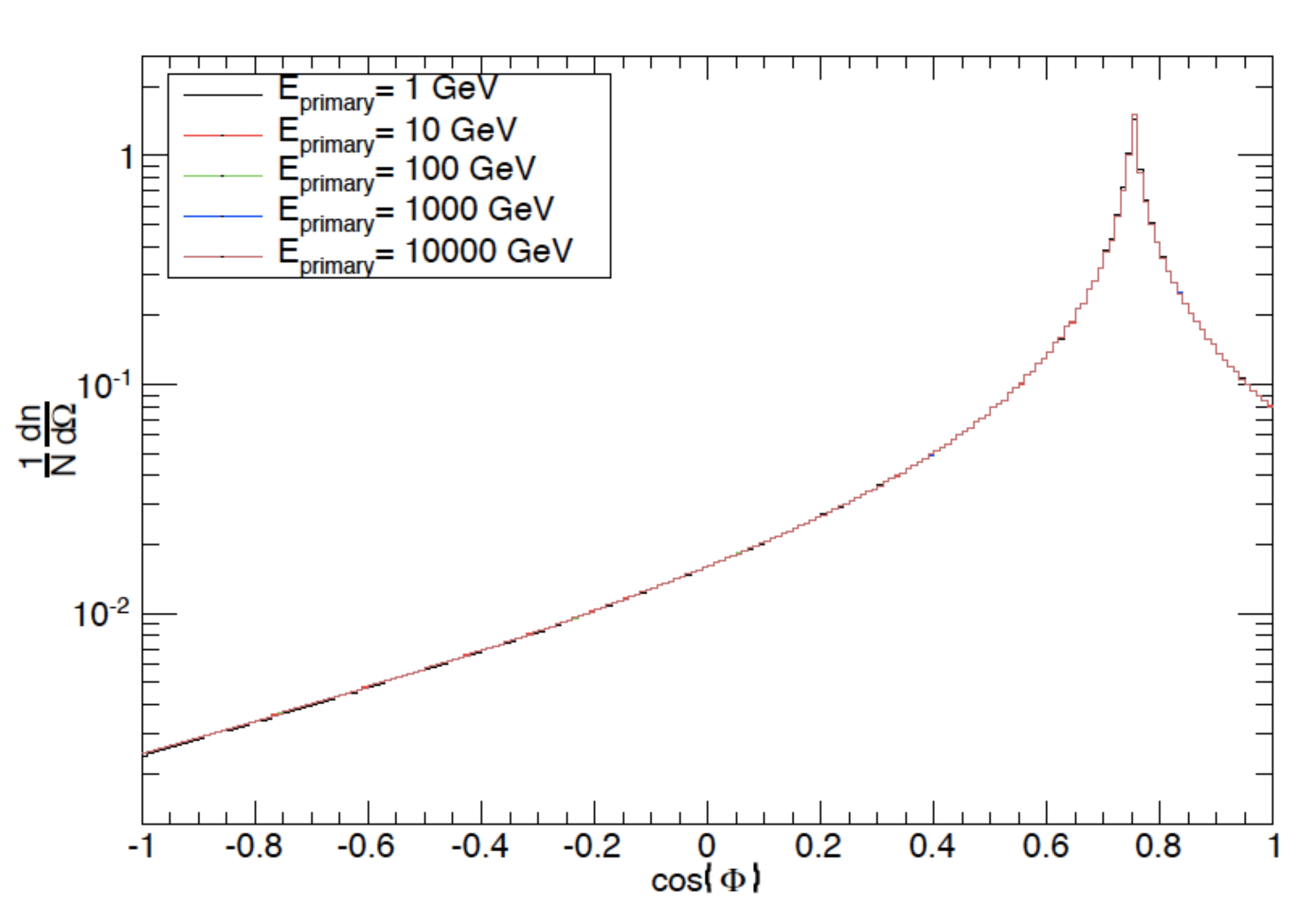}
	\caption{Angular distribution of Cherenkov photons emitted from an electron via an electromagnetic shower in ice. The distribution is normalised per photon and per steradian. The angle $\Phi$ is measured relative to the direction of the primary electron. The distributions for different energies are very similar and overlay each other. Taken from~\cite{raed13}}
	\label{fig-angular}
\end{figure}

The particles within the shower follow an angular distribution which is determined by the bremsstrahlung and pair production kinematics and by scattering of electrons and positrons from the electrons in the medium. According to this angular distribution the angular distribution of Cherenkov photons emitted from electrons and positrons in the shower is broadened with respect to the primary electron direction. Fig.~\ref{fig-angular} shows the angular distribution of Cherenkov photons for electrons moving in ice. The shape of the distribution is almost independent of the electron energy and is almost identical for electrons moving in water. For details on the energy loss mechanisms of electrons see~\cite{patr16}.

\subsection{Signature of muons}

Muons with energies up to several hundred GeV are minimally ionising particles travelling almost perfectly straight through the target medium with a long track length. As an example, the muon track extends to $1\unitkm$ length for a $350\unitGeV$ muon in water. Cherenkov photons are emitted under the Cherenkov angle with an almost uniform emission density along the track. The energy loss per unit track length is almost independent from the muon energy. 

For energies above a TeV the muon energy loss starts to be dominated by stochastic processes---often called stochastic energy losses---like bremsstrahlung, pair production and photonuclear interactions. The produced bremsstrahlung photon and electron-positron pairs induce an electromagnetic shower with its characteristic short shower length, high energy deposition and bright Cherenkov photon emission. As a consequence, for energies above some TeV the energy loss per unit length of the muon is proportional to its energy. For further details see~\cite{patr16}.

\subsection{Signature of taus}

Tau leptons have a lifetime of $\uptau = 0.3\unitps$ only. Accordingly, the tau travel length\footnote{The travel length $l$ is given by $l = c\,\tau\,E/m $ with $c$ the velocity of light, $\tau$ the life time, $E$ the energy and $m$ the mass.}
is $5\unitmm$ for a tau energy of $100\unitGeV$ and $50\unitm$ for a tau energy of $1\unitPeV$. After its travel the tau lepton decays either into a charged lepton plus neutrinos or a quark pair plus neutrino thus leading to the signature of either an electromagnetic shower in case of an electron, a muon track in case of a muon or a hadronic shower in case of quarks, see Fig.~\ref{fig-interaction}. Due to this decay scheme and due to the relatively short travel length of the tau lepton, the signature of a tau-neutrino $\upnu_{\uptau}$ in the telescope mixes with the signatures of the other two neutrino species $\upnu_{\text{e}}$ and $\upnu_{\upmu}$.

\subsection{Signature of hadrons}

The energy which is transferred by the W-boson or Z-boson to the interacting nucleon leads to secondary hadrons which interact with target nucleons thus producing a shower of hadrons. The distribution of transverse momenta in hadronic interactions lead to a slightly broader angular distribution of Cherenkov photons from a hadronic shower compared to an electromagnetic shower. Further, for the same primary energy the amount of emitted Cherenkov photons is lower and the fluctuation in the number of photons is larger for hadronic showers than for electromagnetic showers~\cite{patr16}.

\section{Existing neutrino telescopes}\label{ra_sec4}

For the first neutrino telescopes, neutrino detectors have been built as instruments using a dedicated target material placed in a dedicated housing. With the Kamiokande and Superkamiokande detectors~\cite{iked82,fuku03} using water as target material it was possible to detect rings of Cherenkov light and reconstruct the energy and direction of neutrinos from this signature. The directional sensitivity is the reason to name them a telescope. One huge success was the proof of the directional flux of neutrinos from the sun~\cite{fuku98}.

The need for a large detector mass lead to the idea to instrument a target material in its natural environment. This idea goes back to Markov who formulated it already in 1960~\cite{mark60}. The first attempt to realise this concept was made many years later with the DUMAND detector~\cite{babs90,robe92}, which was located in the deep sea close to Hawaii. It unfortunately failed due to technological problems. It was followed by the BAIKAL detector in the water of Lake Baikal~\cite{belo97}, the AMANDA detector deployed in the glacial ice at the South Pole~\cite{andr00} and the ANTARES detector in the Mediterranean Sea~\cite{ager11}. Based on the success of these projects the larger IceCube detector was built at the South Pole \cite{aart17}, KM3NeT is being built in the Mediterranean Sea~\cite{adri16} and Baikal GVD is under construction in Lake Baikal~\cite{avro16}.

In the following, the largest running telescope in the Southern hemisphere, i.e.~the IceCube detector, and the largest telescope in the Northern hemisphere, i.e.~the ANTARES detector, will be described a bit more in detail.

\begin{figure}[t]
	\centering
	\includegraphics[width=0.99\textwidth]{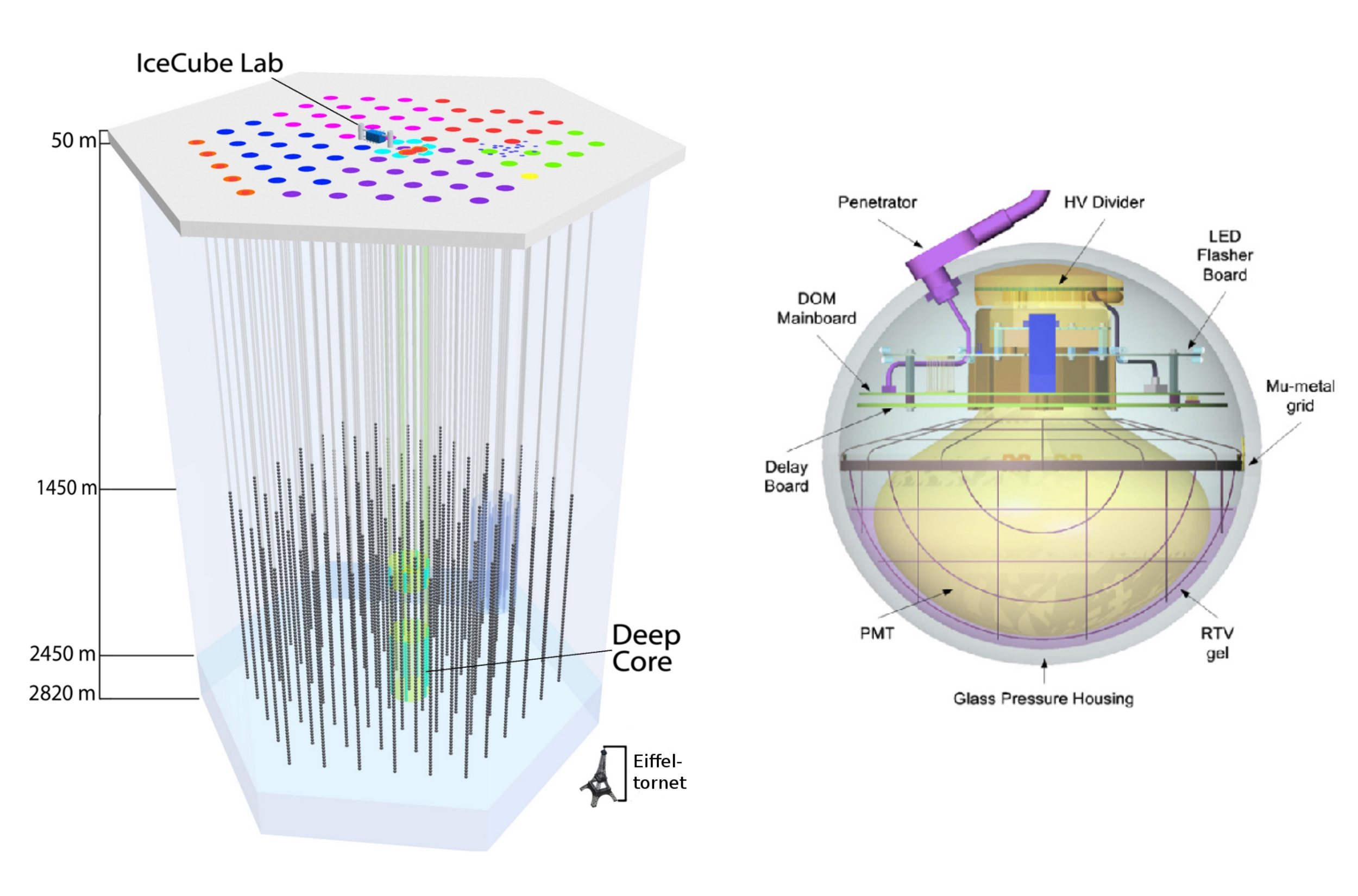}
	\caption{Left: schematic view of the IceCube detector. Black dots indicate the positions of optical modules along the strings. Right: IceCube digital optical module (DOM) housing a 10-inch photomultiplier (PMT). Credit: IceCube Collaboration}
	\label{fig-icecube}
\end{figure}

A schematic view of the IceCube detector is shown in Fig.~\ref{fig-icecube}. The glacier at the South Pole has a thickness of about $3\unitkm$. Strings are deployed in the ice vertically down to $2500\unitm$ below the glacier surface. In a depth between $1450\unitm$ and $2450\unitm$, $60$ optical modules are attached to each string, each module housing a $10$-inch photomultiplier tube. IceCube comprises a total of $86$ strings with a total of $5160$ optical modules. The horizontal distance between strings is $125\unitm$ and the vertical distance between optical modules is $17\unitm$ resulting in an instrumented volume of about $1$ cubic kilometre. According to the distance between strings the energy threshold for the detection of neutrinos is about $200\unitGeV$. A more densely instrumented region which is located in the clearest ice almost in the centre of IceCube, is called "Deep Core". It allows for the detection of neutrinos with a lower energy threshold of about $10\unitGeV$. Further, "IceTop" consisting of $80$ pairs of water Cherenkov tanks with about $2\unitmcb$ water in each tank, complements IceCube with a cosmic ray detector on the glacier surface. IceCube has been completed in 2010 and is taking data continuously.

\begin{figure}[t]
	\centering
	\includegraphics[width=0.82\textwidth]{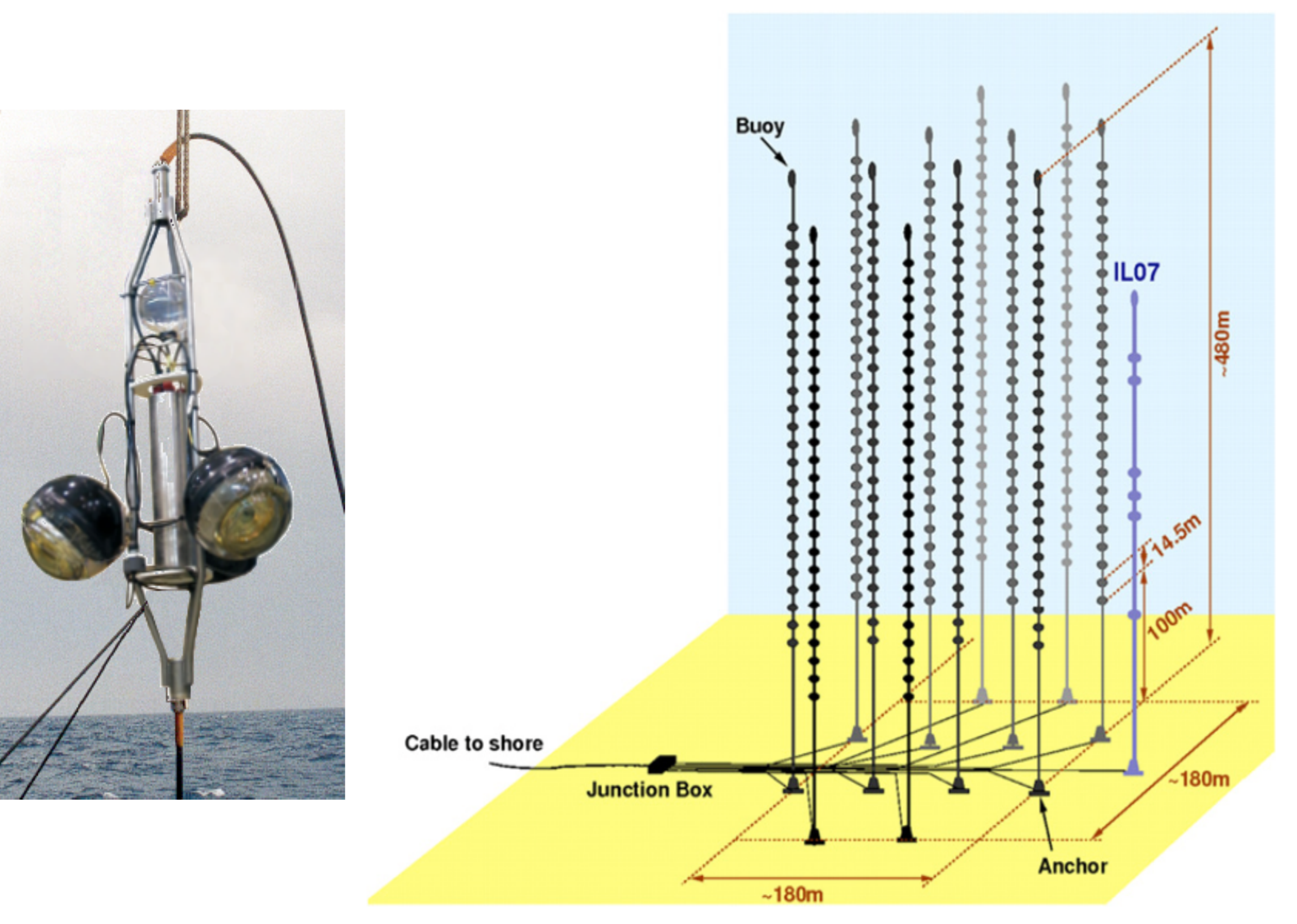}
	\caption{Right: Schematic view of the ANTARES detector. A dot on a string represents an optical storey. Left: A storey comprising 3 optical modules housing a $10$-inch photomultiplier each. Credit: ANTARES Collaboration.}
	\label{fig-antares}
\end{figure}

A schematic view of the ANTARES detector is shown in Fig.~\ref{fig-antares}. ANTARES comprises 12 strings deployed in the sea in $2500\unitm$ depth, $25\unitkm$ off the coast near Toulon, France. Each string extends to $480\unitm$ height from the sea floor and is equipped with optical modules starting $100\unitm$ above the floor. A total of $885$ optical modules are attached to the $12$ strings covering an almost cylindrical volume of $200\unitm$ in diameter and $350\unitm$ in height. An optical module contains a $10$-inch photomultiplier housed in a $17$-inch glass sphere. The spheres are grouped to form triplets integrated in so-called storeys, see Fig.~\ref{fig-antares}. The ANTARES strings are held almost vertical by a buoy at the top of each string. The strings are floating smoothly in the sea current and the position of each optical module is monitored to a precision of about $10\unitcm$ at each instant in time. ANTARES has been completed in 2008 and is taking data continuously.

\subsection{Telescope features and quality parameters}

An important parameter describing the sensitivity of an astronomical telescope is the light collection area, which is essentially determined by the diameter of the telescope mirror. The analogue parameter for a neutrino telescope is the so-called neutrino effective area. It is defined as the size of the hypothetical area $A_{\text{eff}}$ on which $100\,\%$ of neutrinos would be detected when crossing this area. Accordingly, for a neutrino flux $\Phi(E,\Omega)$ the number $N$ of detected neutrinos per time is given by:

\begin{equation}
	N = \int \Phi(E,\Omega)\ A_{\mathrm{eff}}(E,\Omega)\ \text{d}E \text{d}\Omega
\end{equation}

\begin{figure}[t]
	\centering
	\includegraphics[width=0.99\textwidth]{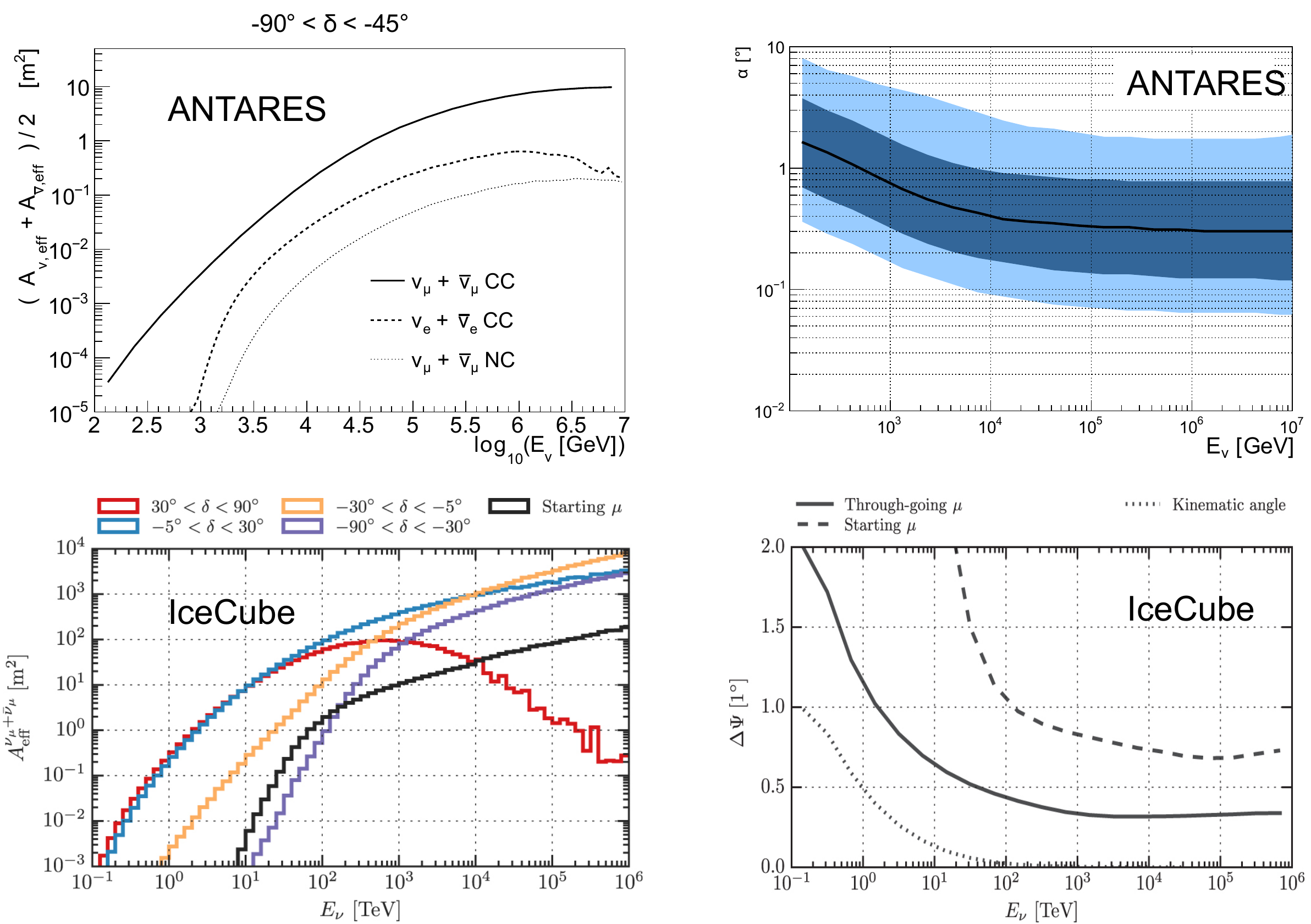}
	\caption{Neutrino effective area (left) and angular resolution (right) for muon neutrinos as function of the neutrino energy for the ANTARES (top) and IceCube (bottom) telescope. $\delta$ denotes the declination in the equatorial coordinate system; $\delta = 90^\circ$ is the direction from North to South. (Taken from~\cite{albe17} (top left), \cite{kouc17} (top right) and from \cite{aart18} (bottom)).}
	\label{fig-eff-area-angular}
\end{figure}

Due to the small neutrino cross section this area is much smaller than given by the diameter of the detector, see Fig.~\ref{fig-eff-area-angular} for ANTARES. Typical values are in the order of a square meter. As the neutrino cross section increases with energy the neutrino effective area increases for energies up to several hundred TeV. The effective area decreases for higher energies for neutrinos coming from below the detector because the Earth starts to be opaque, see Fig.~\ref{fig-eff-area-angular}.

A further important parameter is the angular resolution of the telescope. It is important for the sensitivity to detect point sources and for the suppression of background. As neutrinos are not detected directly but only via the secondary particles produced in neutrino interactions the angular resolution is determined by two effects: first, the kinematic angle between the primary neutrino and the secondary particles and second, the reconstruction quality of the direction and energy of the secondary particles. Both effects depend on the energy of the involved particles leading to an improved angular resolution with increasing energy, see Fig.~\ref{fig-eff-area-angular}. The presented resolution is given for events containing muons which---due to their long tracks---produce an elongated distribution of detected Cherenkov photons, see Fig.~\ref{fig-events}. 

\begin{figure}[t]
	\centering
	\includegraphics[width=0.90\textwidth]{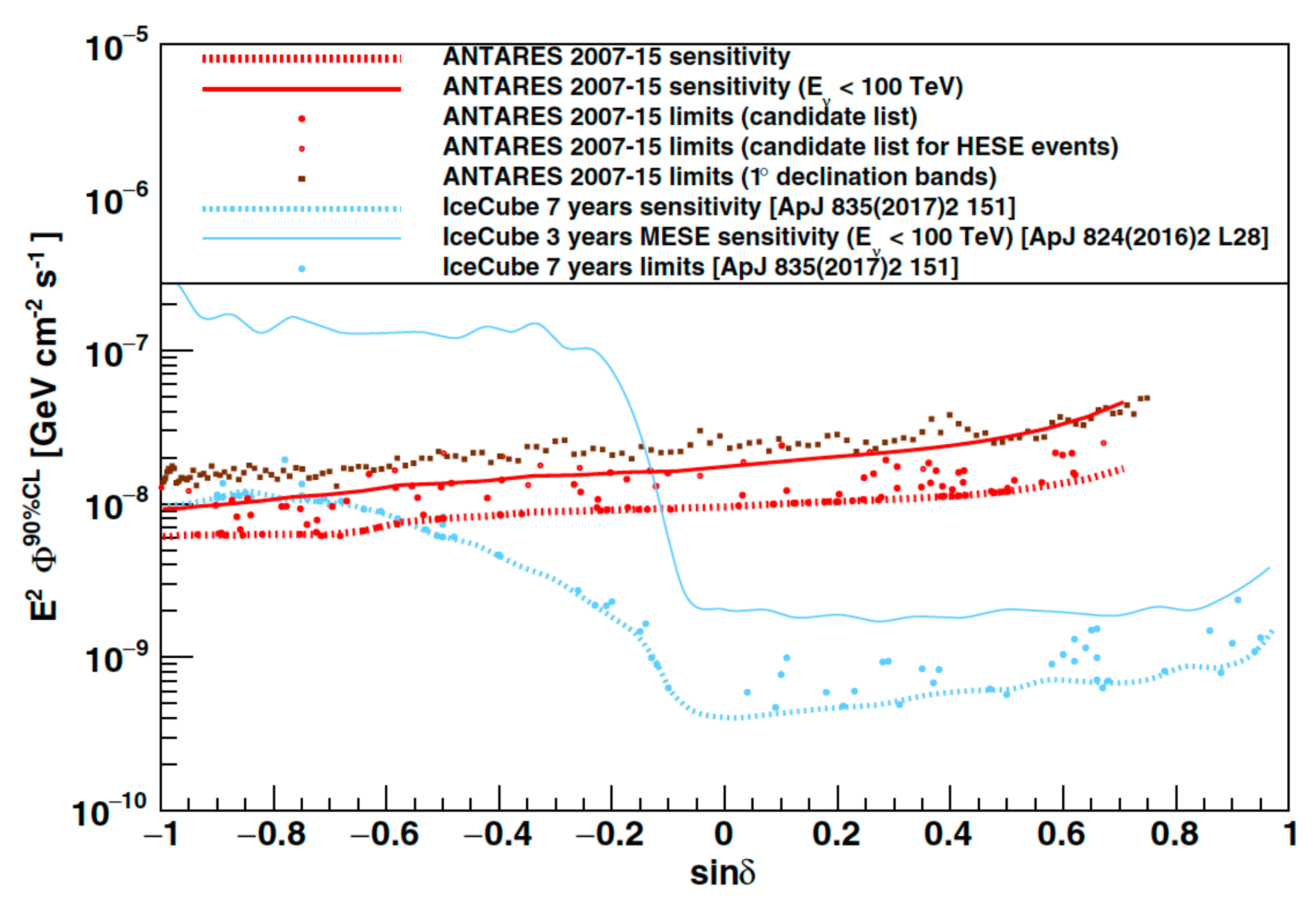}
	\caption{Sensitivity and upper limits to the neutrino flux from point sources as function of the declination angle $\delta$ assuming an $E^{-2}$ spectrum. The Flux in units of particles per square centimeter, seconds and GeV is multiplied by energy squared. HESE (MESE): High (medium) energy starting events. (taken from~\cite{albe17})}
	\label{fig-sensitivity}
\end{figure}

Based on angular resolution and effective area, the sensitivity of the telescope to neutrino point sources can be determined. This sensitivity is presented in Fig.~\ref{fig-sensitivity}. It is defined as the average of the upper limit on an assumed signal flux that can be deduced from background according to the Neyman method~\cite{neym37}\cite{feld98}. The astrophysical neutrino flux sensitivity is limited by the background of diffuse atmospheric neutrinos. As no point sources have been discovered so far, upper limits on the source flux can be derived from the data. Measured upper limits with $90\,\%$ confidence level on the neutrino point source flux assuming an $E^{-2}$ energy dependence are shown in Fig.~\ref{fig-sensitivity}.
Another important parameter is the energy of the neutrino. The amount and geometrical distribution of detected Cherenkov photons deliver a measure of the energy of the secondary particles and from this information the energy of the primary neutrino is estimated. Due to the involved fluctuations and due to the low density of photo sensors the energy resolution is rather weak with values of about $0.1$ to $1$ units in $\log(E)$.

As is the case for most detection instruments also neutrino telescopes are affected by backgrounds. Some of these effects will be shortly described in the following.

On the one hand the detection of muons is advantageous due to their long tracks which are detectable even if the muon is produced far outside the detector. This effect leads to a large effective area for muon-neutrinos. On the other hand, muons which are produced in cosmic ray interactions in the Earth atmosphere---so-called atmospheric muons---can reach the detector and can thus mimic neutrino events. One possibility to suppress the atmospheric muon background is the selection of muons which arrive from below the detector and produce up-going tracks. Only neutrinos can travel through the Earth so that this directional cut selects neutrinos having produced the detected muon. The need to suppress atmospheric muons leads to an angular acceptance of the neutrino telescope covering predominantly the hemisphere below the detector. Accordingly, the IceCube telescope at the South Pole can observe neutrino sources in the Northern hemisphere while the ANTARES telescope is largely complementarily sensitive to sources in the Southern hemisphere. Due to its large volume IceCube on average is far more sensitive than the ANTARES telescope. Nevertheless, for the southern hemisphere of the sky---which is the direction of up-going neutrinos for a telescope on the northern hemisphere---ANTARES reaches a similar sensitivity as IceCube, see Fig.~\ref{fig-sensitivity}.

An opportunity to actively suppress the atmospheric cosmic ray background is the selection of so-called starting events. The distribution of detected Cherenkov photons can be analysed to reconstruct the starting point from which secondary particles leave the neutrino interaction vertex. The selection of events starting inside the detector volume excludes atmospheric muons. It selects neutrinos including those from cosmic ray interactions in the atmosphere. If wanted, some of these atmospheric neutrinos can be rejected via the detection and vetoing of an accompanying muon which is produced in the same primary cosmic ray interaction and reaches down to the detector. The selection of starting events considerably reduces the effective area of the telescope but opens its sensitivity to down-going neutrinos.

\subsection{Water and ice as detector media}

Water and ice are similar detector media. As can be deduced from formula (1), the slightly higher electron density of water compared to ice leads to a higher refractive index $n_{\text{water}} = 1.34$ resp.~$n_{\text{ice}} = 1.32$\,\footnote{at the wavelength of $400\unitnm$} and thus a $5\,\%$ enhanced Cherenkov photon emission density in water. Further, it leads to slightly different Cherenkov emission angles of $42\unitdeg$ in water and $41\unitdeg$ in ice.

In addition to the production of Cherenkov photons the conditions for the propagation of these photons from the emission point to the optical sensors are important. The absorption length $ \lambda_{\text{abs}} $ parametrises the propagation length of photons in a medium before being absorbed (1/e-probability to survive after path length $ \lambda_{\text{abs}}$). The effective scattering length $\lambda_{\text{scat}} $
parametrises the angular scattering of photons propagating through the medium. For the water at the ANTARES site, the absorption length is $ \lambda_{\text{abs}}^{\mathrm{ANTARES}} = 55\unitm$ and the effective scattering length is $\lambda_{\text{scat}}^{\text{ANTARES}} = 265\unitm$ for a photon wavelength of $470\unitnm$~\cite{agui05}. These parameters are constant over the volume of the detector.

\begin{figure}[t]
	\centering
	\includegraphics[width=0.90\textwidth]{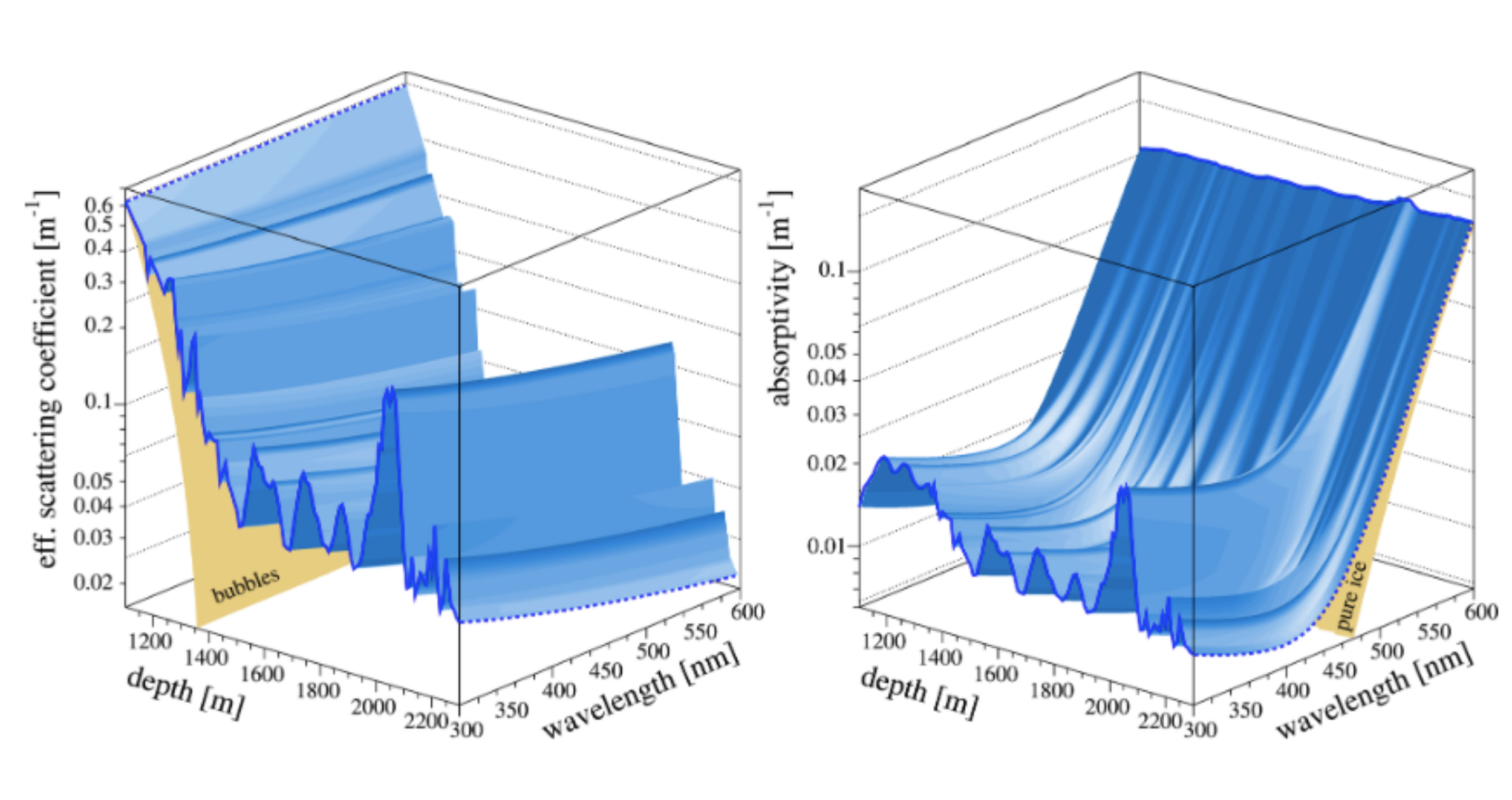}
	\caption{Effective scattering length = 1/eff.~scattering coefficient (left) and absorption length = 1/absorptivity (right) of the glacial ice at the South Pole (taken from~\cite{acke06})}
	\label{fig-icelayers}
\end{figure}

For the ice at the IceCube site the absorption length $\lambda_{\text{abs}}^{\text{IceCube}}$ and scattering length $ \lambda_{\text{scat}}^{\text{IceCube}}$ vary considerably for different positions in the detector due to the evolution of the layer structure of the glacial ice, see Fig.~\ref{fig-icelayers}. Typical values for deep clear ice are $\lambda_{\text{abs}}^{\text{IceCube}} = 110\unitm$ and $\lambda_{\text{scat}}^{\text{IceCube}} = 30\unitm$~\cite{acke06}. As can be seen from Fig.~\ref{fig-icelayers}, a so-called dust layer with very short absorption length appears at a depth from $1950\unitm$ to $2100\unitm$. The absorption length and the scattering length have to be known as well as possible in order to take the influence of these parameters into account for the reconstruction of particle direction and energy from the detected Cherenkov photons. Due to the inhomogeneity of the ice, the parameters strongly vary for different positions and direction of particles propagating through the detector. The IceCube collaboration is performing measurements and analyses improving the knowledge of the ice properties continuously. Since the ice is stable over time, this improvement allows to improve the reconstruction quality of all detected events during the lifetime of the detector. 

Another important parameter is the optical noise. While an IceCube photomultiplier in the glacial ice sees a rather low and timely stable noise rate of below $1\unitkHz$, an ANTARES photomultiplier sees a rate of at least $40\unitkHz$ going up to several $100\unitkHz$ rather often. The ANTARES photomultiplier rate is due to two effects: first, there is an almost constant rate caused by the radioactive decay of Potassium-40 (\isotope[40]{K}) nuclei in the sea water and the related Cherenkov photon emission from the \isotope[40]{K} decay electrons. Second, there is a strongly variable and season-dependent contribution from bioluminescent light emission. In order to handle the issue ANTARES monitors the rates every $0.1$ second and detects photons with a nanosecond time resolution. Fortunately, a neutrino event signature causes a time correlation of detected Cherenkov photons in a time residual window\footnote{The time residual is the difference of measured photon detection time and expected photon detection time for a given event hypothesis.} of a few nanoseconds during which the contribution from the optical noise is almost negligible.

\begin{figure}[p]
	\centering
	\includegraphics[width=0.8\textwidth]{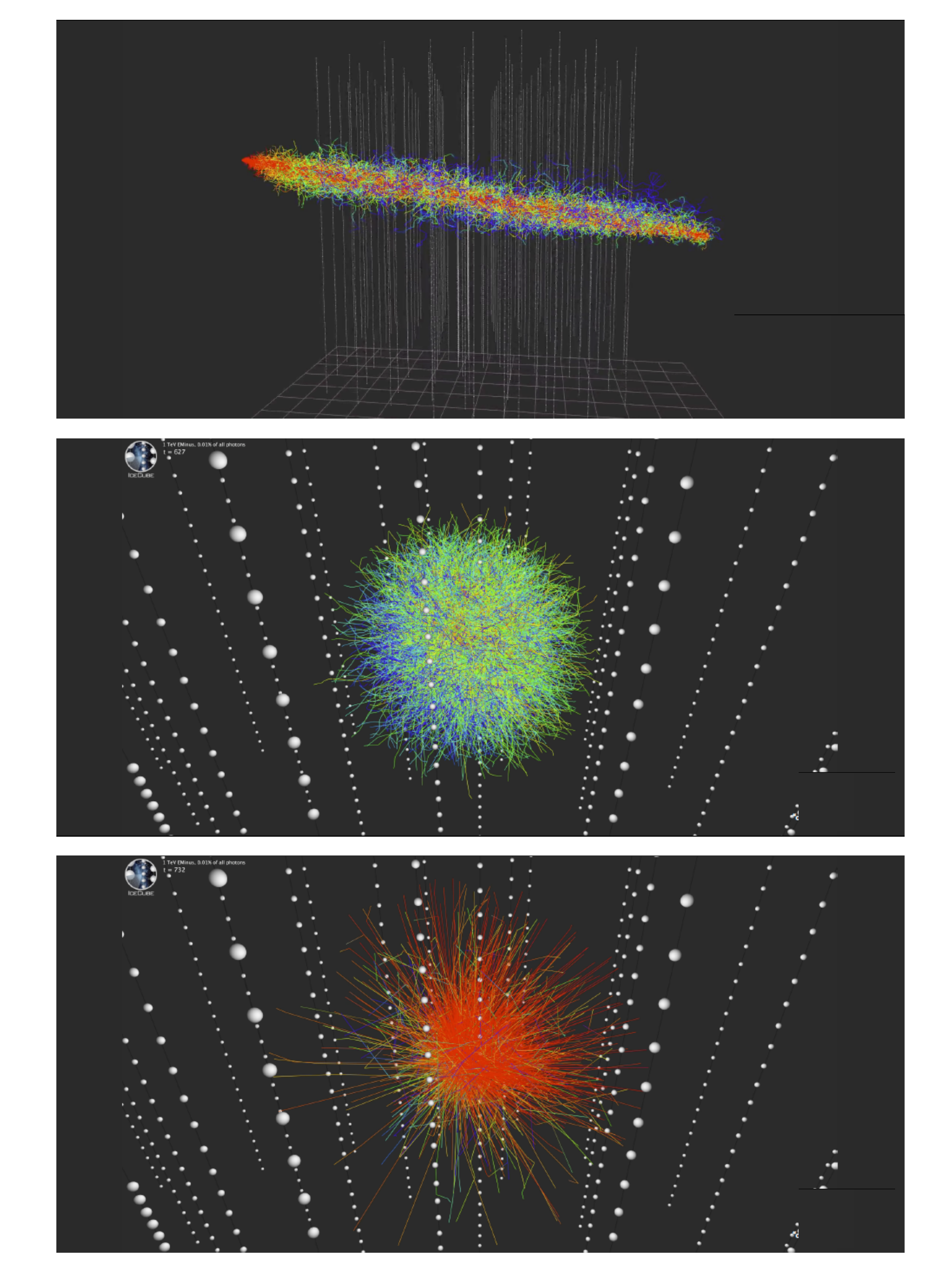}
	\caption{Visualisation of the Cherenkov photon emission and the photon paths in a detector medium. The paths of single photons are drawn as a thin lines including directional changes due to scattering. The colour encodes the time to which the single photon is delayed relative to the time of a straight path (red: on-time, blue: delayed time). Top image: muon with $100\unitTeV$ energy in South Pole ice, mid: electromagnetic shower of $1\unitTeV$ energy in South Pole ice, bottom: electromagnetic shower of $1\unitTeV$ energy in ANTARES water with IceCube detector structure. Only $0.01\,\%$ of all Cherenkov photons are shown in the plots (taken from~\cite{kopp15}).}
	\label{fig-events}
\end{figure}

The influence of the properties of water resp.~ice is illustrated in Fig.~\ref{fig-events}. As a consequence of the differences of the absorption length and the scattering length in ice and water one can say that a photon in IceCube travels with many scatterings before being absorbed and a photon in ANTARES travels a shorter pathlength but effectively without scattering. These effects are visualised for three example events. Fig.~\ref{fig-events} shows in the top a muon of $100\unitTeV$ energy propagating through IceCube. The muon emits Cherenkov photons with almost constant emission density along its track. Thin lines show the path of single Cherenkov photons. The colour encodes the time of a photon relative to the time that this photon had taken if it would have propagated on a straight line. Obviously, the time delay increases with increasing photon path length due to scattering. The image in the middle of Fig.~\ref{fig-events} shows an electron of $1\unitTeV$ energy travelling through IceCube. The electron induces an electromagnetic shower that deposits its energy almost instantaneously thus appearing as a point-like source of Cherenkov photons. The photons seem to be emitted almost isotropically. Only a minor signature of forward emission can be seen in the less scattered (green compared to red) photons. The image in the bottom of Fig.~\ref{fig-events} shows the signature of an electron of the same energy in the ANTARES detector. Compared to IceCube, less photons arrive 
at the same distance to the shower but these photons travelled almost straight keeping the information of the direction of the primary electron and the related Cherenkov angular emission quite accurately. 

\begin{figure}[t]
	\centering
	\includegraphics[width=0.95\textwidth]{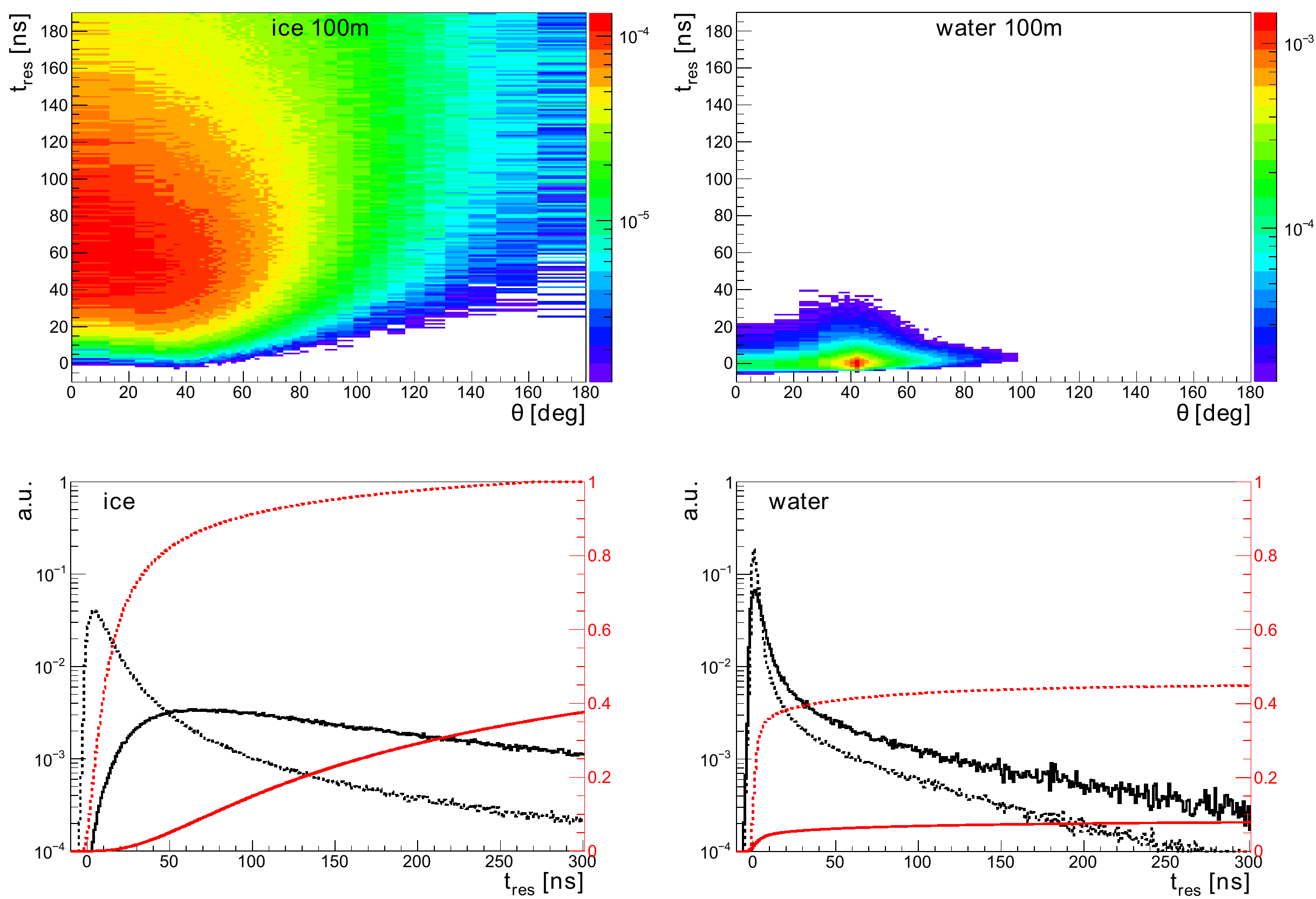}
	\caption{Top: Angular distribution versus time residual for Cherenkov photons from an electromagnetic shower in ice (top left) and water (top right) for $100\unitm$ distance from the shower vertex. Bottom: time residual $t_{\text{res}}$ distribution (black line) and integrated time residual distribution (red line) of Cherenkov photons from an electromagnetic shower in $30\unitm$ (dotted line) resp.~$100\unitm$ (full line) distance from the shower vertex\tablefootnote{Photons are counted every time when transversing the $30\unitm$/$100\unitm$ shell, so that individual photons can be counted multiple times.} in ice (bottom left) and water(bottom right)\tablefootnote{The plots in Fig.~\ref{fig-ice-water} are calculated with the parameters $\lambda_{\text{abs}}^{\text{water}} = 55\unitm$, $\lambda_{\text{abs}}^{\text{ice}} = 110\unitm$, $\lambda_{\text{scat}}^{\text{water}} = 250\unitm$, $\lambda_{\text{scat}}^{\text{ice}} = 30\unitm$.}. (taken from~\cite{jann17})}
	\label{fig-ice-water}
\end{figure}

The influence of the target medium parameters are displayed in more detail in Fig.~\ref{fig-ice-water}. With increasing distance from the emission point the angle of a Cherenkov photon (defined as the angle between the direction of the primary electron and the direction given by the line from the shower vertex to the actual photon position) is increasingly smeared out by scattering in the medium. Further, the time at which the photon arrives at a certain distance from the emission point is delayed. The related time residual $t_{\text{res}}$ is defined as the difference between the detected time and the expected time assuming a photon travel on a straight path, see Fig.~\ref{fig-ice-water}. Angular and time smearing deteriorate the precision to which the direction of the incident particle can be reconstructed. The effect is considerably stronger for ice than for water resulting in a better angular resolution of particles in Mediterranean Sea water than in South Pole ice. Accordingly, the direction of shower events can be reconstructed much better in water than in ice. For muons the angular resolution of ANTARES is only slightly better than that of IceCube (see Fig.~\ref{fig-eff-area-angular}) because IceCube is much larger than ANTARES and thus can detect Cherenkov photons from a long part of the muon track delivering a large lever arm for angular reconstruction, 
see also the top event of Fig.~\ref{fig-events}. As a drawback for water, less photons survive for a given distance. Especially when aiming at the highest neutrino energies (above $1\unitPeV$) and the related largest neutrino telescope, ice may be superior to water because in order to detect a reasonable number of Cherenkov photons it needs much less dense photo sensor instrumentation. On the other hand, scattering in ice distorts the information carried by the photons. It should become obvious from this information that the design of a neutrino telescope is different in ice and in water. 

\section{Future neutrino telescopes}\label{ra_sec5}

IceCube has discovered a diffuse cosmic neutrino flux in 2013~\cite{aart13}. Despite this important success the existing neutrino telescopes are still too small to investigate this diffuse flux in the energy regime above $1\unitPeV$ and too small to detect astrophysical point sources. Further, a low energy neutrino telescope would allow a more precise measurement of the oscillation signature of atmospheric neutrinos with sensitivity to the neutrino mass ordering~\cite{perez20}. Accordingly, there are ongoing activities aiming for the next generation of neutrino telescopes.

In lake Baikal the new detector Baikal GVD (Gigaton Volume Detector)~\cite{avro16} aiming at the $1\unitTeV$ to $10^6\unitTeV$ neutrino energy regime, is in the first phase of construction with two out of the planned $8$ clusters of strings being deployed. The first phase will cover a volume of $0.4\unitkmcb$, the final size with $27$ clusters should reach a volume of $1.5\unitkmcb$.

\begin{table}[t]	
\vspace{-3mm}
\end{table}

IceCube is considering a two-fold extension called IceCube Gen2~\cite{sant17}. One part is dedicated to the highest neutrino energies with a detector of about $10$ cubic kilometers instrumented volume. Due to the envisaged energy regime the optical modules will be distributed more sparsely, i.e.~with $250\unitm$ to $300\unitm$ horizontal distance between strings. The other part---named PINGU (Precision IceCube Next Generation Upgrade)~\cite{aart17a}---is dedicated to lower the energy threshold down to about $1\unitGeV$ in order to obtain sensitivity to oscillation signatures and the neutrino mass ordering in the atmospheric neutrino flux. 

KM3NeT has published its letter of intent in 2016~\cite{adri16}. This project comprises the low energy detector KM3NeT/ORCA close to the ANTARES site near Toulon, France, and the detector KM3NeT/ARCA near Capo Passero in Sicily, Italy. ORCA will use a densely instrumented cylindrical volume of about $200\unitm$ in diameter and $150\unitm$ in height ($5$ Megaton) with a total of $2070$ optical modules each comprising $31$ $3$-inch photomultipliers. ORCA is designed to measure the oscillation signatures of atmospheric neutrinos in the energy range from $1$ to $30\unitGeV$ with the main aim to determine the neutrino mass ordering. KM3NeT/ARCA will consist of two building blocks, each of a cylindrical volume with a diameter of $\unitkm$ and a height of $600\unitm$ equipped with $2070$ optical modules on $115$ strings. ARCA is primarily designed for the search of galactic neutrino point sources. ARCA and ORCA are using the same technology with the only difference of the distance of optical modules along strings and between strings. KM3NeT has received funding for the phase-1 of its project and is deploying the first $10$ percent of the full detector.

\cfoot{\thepage}														
\rehead{}																	
\lohead{}																	
\rohead{}																	
\lehead{}																	

\bibliography{AntonG-Neutrino_Telescopes}

\clearpage

\end{document}